\documentclass[10pt,pdftex]{article}
\usepackage{mathptmx}
\usepackage{indentfirst}
\usepackage{lipsum}
\usepackage{algpseudocode}
\usepackage{graphicx}
\usepackage{float}
\usepackage[most]{tcolorbox}
\usepackage{amsmath, amssymb, amsthm}
\usepackage{xcolor}
\usepackage{bm}
\usepackage{tikz}
\usepackage{float}
\usepackage[width=.7 \textwidth]{caption}
\usepackage[ruled, vlined]{algorithm2e}
\usepackage{multirow}
\usepackage{caption}
\usepackage{tcolorbox}

\begin{document}

\begin{center}
\begin{large}

Bayesian Statistics: A Review and a Reminder for the Practicing Reliability Engineer

\end{large}

Carsten H. Botts

The Johns Hopkins University Applied Physics Lab

Laurel, MD 20723

\end{center}

\abstract{This paper introduces and reviews some of the principles and methods used in Bayesian reliability. It specifically discusses methods used in the analysis of success/no-success data and then reminds the reader of a simple  Monte Carlo algorithm that can be used to calculate the posterior distribution of a system's reliability. This algorithm is especially useful when a system's reliability is modeled through the reliability of its subcomponents, yet only system-level data is available.}

\vspace{.5cm}

\noindent {\bf Keywords}: Success/No-Success Data, Monte Carlo Methods, System Reliability











\section{Introduction}

A common reliability metric of a system is the probability that it will pass, or survive, a stress test.  Several tests of the system are typically necessary to learn about such a probability, but conducting many tests of a sophisticated system may be prohibitively expensive.  Bayesian statistical methods can help in such a situation, since they make it possible for one to include other types of data (such as computer simulation experiments or the opinion of a subject-matter expert) into the  statistical analysis. Bayesian methods may also be necessary because many modern systems do not fail during testing. With no failures, it is difficult for classical statistics to accurately quantify the probability of failure. 

This paper first provides a brief and general review of Bayesian methods. Section \ref{sctn:BayesBernoulli} then discusses how these methods can be used to learn more about the probability of a system surviving a  test. This section concludes by reminding the reader of a straight-forward  algorithm for calculating a total system's reliability once it has been tested. This algorithm is simple, produces an exact answer, and is rarely mentioned in the Bayesian reliability literature.

\section{Bayesian methods}

In this section we introduce the basics of Bayesian principles and Bayesian statistical methodology. The most effective way to introduce this concept  is to contrast it with the principles and methodology of classical statistics. The biggest difference between Bayesian and classical statistics is with regard to how probability is defined.     In classical statistics, probability is the long-run frequency of an event. So for a fixed (and unknown) parameter such as a population mean, $\mu$, \begin{equation}\label{eqn:FreqProb} {\mathbb P} \left( 3.66 \leq \mu \leq 4.11 \right) = \left \{ \begin{array}{ll} 1 & {\rm if~true} \\ 0 & {\rm if~not}  \end{array} \right. .\end{equation} In words, Equation (\ref{eqn:FreqProb}) states that the fixed parameter $\mu$ is either in the stated interval or it is not.

Bayesian statisticians think about probability in a different way. In Bayesian statistics, probability is the belief that a statement is true. So if one believes (based on their experience and/or the data that they have seen) that $\mu$ is within the stated interval with 95\% probability, it would be fair to say that \begin{equation}\label{eqn:BayesProb} {\mathbb P} \left( 3.66 \leq \mu \leq 4.11 \right) = .95.\end{equation} The objective and point of Bayesian statistics is to calculate probabilities like the one in Equation (\ref{eqn:BayesProb}), and to assure that this calculation is scientifically respected.

To calculate such a probability, a Bayesian statistician begins with a prior distribution.  Assuming the unknown parameter of interest is $\theta$, this prior distribution is typically denoted as $\pi(\theta)$.   The prior distribution indicates where the user believes the parameter $\theta$ to be before data is observed and/or collected.   Assume, for example, that we purchased a coin at a magic shop. Upon the purchase, the owner of the shop tells us that the coin will more often turn up heads than tails. In this case, we will let $\theta = {\mathbb P}(H)$, and define the prior distribution $\pi(\theta)$  for all values of $\theta$ between 0 and 1. This prior will also be more heavily weighted towards values of 1 to indicate that, a-priori, the coin is expected to turn up heads more so than tails.

Once the prior is formulated, data is collected. The distribution of the data conditioned on a value of $\theta$ is written as $p({\bf x} | \theta )$, i.e., 

$$\left( X_1, X_2, \ldots, X_n \right)  \sim p \left( x_1, x_2, x_3, \ldots, x_n | \theta \right) = p( {\bf x}|\theta),$$ where ${\bf x} = \left( x_1, x_2, \ldots, x_n \right).$  The function $p \left( {\bf x} | \theta \right)$ is also referred to as the likelihood of $\theta$. 

With the prior and the likelihood, the posterior distribution (typically denoted as $\pi(\theta| {\bf x})$) can be calculated.    The posterior is calculated using Bayes' rule (see \cite{Bayes}). This calculation is shown below.  \begin{equation} \label{eqn:BayesRule} \pi \left( \theta | {\bf x} \right) = {\frac{p({\bf x}, \theta)}{p({\bf x})}} = {\frac{p({\bf x}|\theta) \pi(\theta)}{ \displaystyle \int_{\Theta} p({\bf x}|\theta)\pi(\theta)d\theta}} \propto p( {\bf x}|\theta) \pi(\theta), \end{equation}  where $\Theta$ is the set of all possible values of $\theta$.    The formula given in Eqn (\ref{eqn:BayesRule}) makes sense: the posterior is proportional to the prior distribution of $\theta$ (where we thought $\theta$ was before collecting data) times the likelihood (where the data suggests $\theta$ to be).

 In this paper, we focus on how someone can use Bayesian methods to learn more about the probability that a system survives a test.  Subsection \ref{sctn:PriorPostOneSubsystem} specifically discusses how Bayesian methods are used to learn about the survival probability of one system, and Subsection \ref{sctn:PostEntireSystem} discusses the methods necessary to learn about the survival probability of one system composed of multiple subsystems.  The algorithm discussed in Subsection \ref{sctn:PostEntireSystem} is elementary and uses no approximations when calculating its answer.

\section{Bayesian methods for Bernoulli experiments}\label{sctn:BayesBernoulli}

\subsection{The Prior and Posterior of One Subsystem}\label{sctn:PriorPostOneSubsystem}

Let us first assume that we are interested in learning about   the probability that a system passes an endurance test of some sort.   We will denote this probability as $\theta$, and we will conduct $n$ trials/tests on this system and record $X ,$ the number of times (out of the $n$ trials) that it passes a test.

To do a Bayesian analysis on $\theta$, we begin by specifying a prior distribution for it. A prior often used for the probability of success in a sequence of success/failure trials  is the beta distribution (see \cite{Gelman} and \cite{Carlin}). The beta distribution is specified by two parameters and is especially  convenient in cases such as this  since it is a conjugate prior, i.e., it produces a posterior distribution of the same form. The beta prior takes the form  $$\pi \left( \theta \right) = {\frac{\Gamma \left( \alpha + \beta \right)}{\Gamma(\alpha) \Gamma(\beta)}} \theta^{\alpha - 1} \left( 1 - \theta \right)^{\beta - 1}~~~~~~0 \leq \theta \leq 1,$$ where $\Gamma \left( \cdot \right)$  is the gamma function. This prior has mean $${\rm Prior~Mean} = \frac{\alpha}{\alpha + \beta}$$ and variance $${\rm Prior~Var} = {\frac{\alpha \beta}{ \left( \alpha + \beta \right)^2 \left( \alpha + \beta + 1 \right)}}.$$ The values of $\alpha$ and $\beta$ ($\alpha, \beta > 0$) are selected to reflect a user's prior belief.  This prior belief is often informed in a variety of ways: through expert opinion, computer simulation, prior experiments, etc. If one believed that $\theta$ is small (less than 0.5), one would set $\alpha < \beta$ (making the prior mean less than $0.5$).  If one believed that $\theta$ was large (greater than $0.5$), one would set $\alpha > \beta$. The confidence in these prior beliefs is, of course, reflected in the variance of the prior. If one wanted to set the prior mean of $\theta$ to be 0.4, he could set $\alpha = 2$ and $\beta = 3$, making the prior variance 0.04.   If one wanted to elevate the confidence in this statement (that the prior mean of $\theta$ is 0.4), he could adjust the values of $\alpha$ and $\beta$ to $\alpha = 20$ and $\beta = 30$, making the prior variance 0.004.  And if one knew absolutely nothing about $\theta$, he would set $\alpha = \beta = 1$, in which case the prior for $\theta$ is uniform over the interval $(0,1)$.

There is no consistent selection of $\alpha$ and $\beta$ in reliability studies.  Leoni et al. (see \cite{Leoni}) set $\alpha = 3$ and $\beta = 1$ in one of their reliability studies, Burke et al. (see \cite{Burke}) set $\alpha = 7.2$ and $\beta = 0.8$ in one of their reliability studies, and Martz et al. (see \cite{Martz3}) set $\alpha = 27.3$ and $\beta = 0.5$. In those cases where priors are informed by simulation results or previous experiments, analysts sometimes set $$\alpha = n_{\rm pr} \cdot {\hat \theta}^{\rm pr} + 1 ~~~{\rm and}~~~~\beta =  n_{\rm pr} \cdot  \left( 1 - {\hat \theta}^{\rm pr} \right) + 1,$$ where ${\hat \theta}^{\rm pr}$ is a prior estimate of $\theta$, and $n_{\rm pr}$ is some positive number which represents the confidence the analyst has in the simulation or experiment informing the prior (see \cite{Johnson}); one can think of $n_{\rm pr}$ as the effective sample size which informs the prior. The greater this effective sample size which informs the prior, the more peaked the prior distribution is near the prior estimate of $\theta$. If no confidence exists in the simulation informing the prior, then $n_{\rm pr} = 0$ and the prior would be flat.  

\begin{figure}[H]
   \begin{center}
      \includegraphics[height=3in, width = 3in]{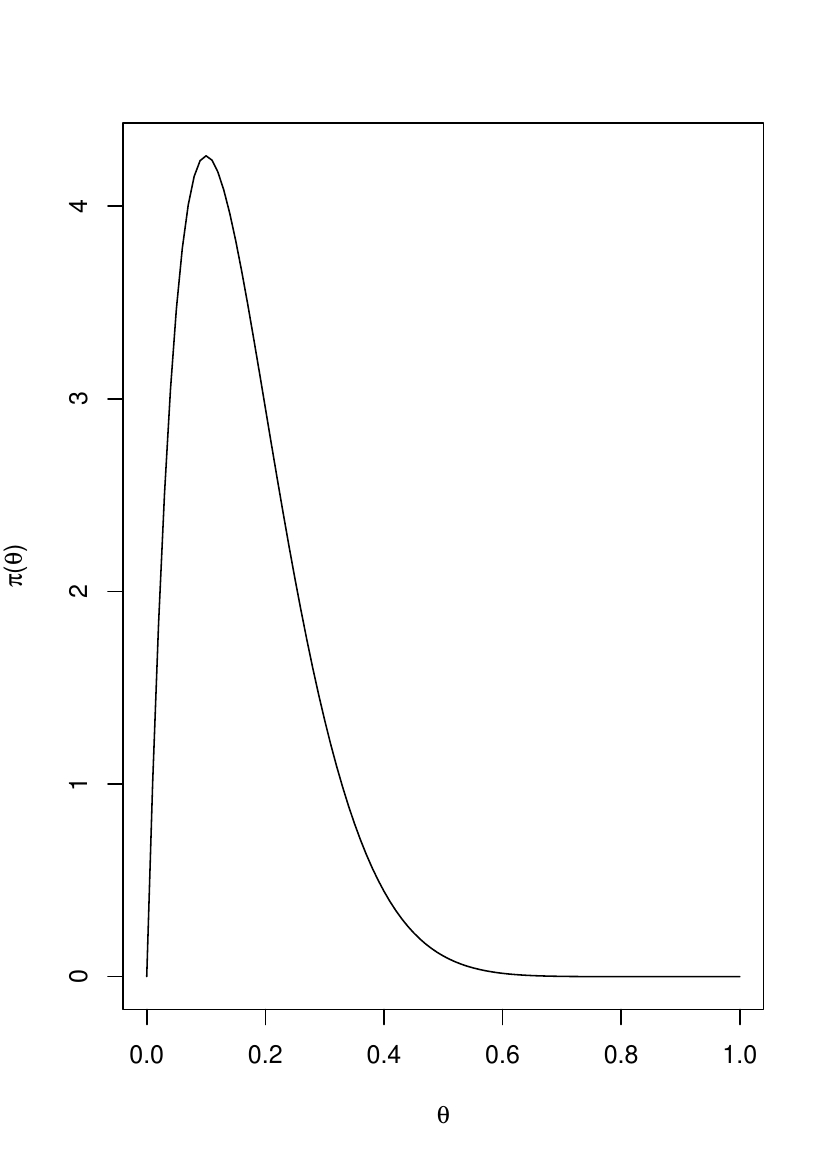}
   \caption[example] 
   { \label{fig:LowProbabilityPrior} The prior $\pi \left( \theta \right)$ with $\alpha = 2$ and $\beta = 10$.}
   \end{center}
\end{figure}

\begin{figure}[H]
   \begin{center}
      \includegraphics[height=3in, width = 3in]{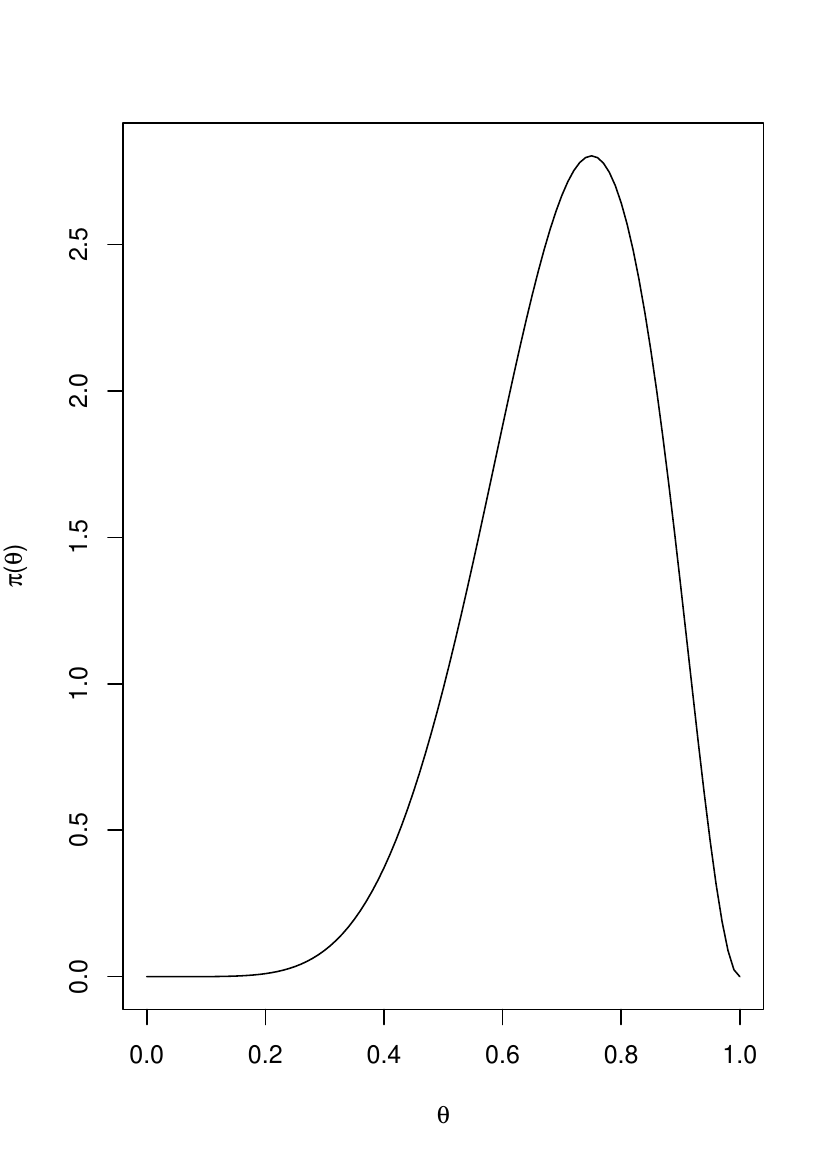}
   \caption[example] 
   { \label{fig:HighProbabilityPrior} The prior $\pi \left( \theta \right)$ with $\alpha = 7$ and $\beta = 3$.}
   \end{center}
\end{figure}

Figures \ref{fig:LowProbabilityPrior} $\textendash$ \ref{fig:HighProbabilityPrior} illustrate what these prior distributions look like. The prior in Figure \ref{fig:LowProbabilityPrior} puts large probability on low values of $\theta$ and does so by setting the value of $\alpha$ to be significantly less than the value of $\beta$. The prior in Figure \ref{fig:HighProbabilityPrior} puts large probability on high values of $\theta$, and does so by setting the value of $\beta$ to be smaller than the value of $\alpha$. Also observe that the prior is much more peaked for small values of $\theta$ in Figure \ref{fig:LowProbabilityPrior} than it is for large values of $\theta$ in Figure \ref{fig:HighProbabilityPrior}.  This is a consequence of the difference in the values between $\alpha$ and $\beta$. The difference is larger for the prior in Figure \ref{fig:LowProbabilityPrior} than it is for the prior in Figure \ref{fig:HighProbabilityPrior}.

Let us now assume we observe  $x$ successes of the system out of $n$ tests conducted. In this case the likelihood is binomial,  $$p \left( x | \theta \right) = {n \choose x} \theta^x \left( 1 - \theta \right)^{n-x},$$ making the posterior distribution of $\theta$  \begin{equation} \label{eqn:BetaPosterior} \pi \left( \theta | x \right) = {\frac{p \left( x | \theta \right) \pi \left( \theta \right)}{ \displaystyle \int_{\Theta} p (x | \theta) \pi \left( \theta \right) d\theta}} = {\frac{ \displaystyle {n \choose x} \theta^x \left( 1 - \theta \right)^{n-x} {\frac{\Gamma(\alpha + \beta)}{\Gamma(\alpha) \Gamma(\beta)}} \theta^{\alpha - 1} (1- \theta)^{\beta - 1}}{ \displaystyle \int_0^1 \left[  {n \choose x} \theta^x \left( 1 - \theta \right)^{n-x} {\frac{\Gamma(\alpha + \beta)}{\Gamma(\alpha) \Gamma(\beta)}} \theta^{\alpha - 1} (1- \theta)^{\beta - 1} \right] d \theta}}.\end{equation}  There is a simple way to calculate the posterior distribution in Equation (\ref{eqn:BetaPosterior}) without having to evaluate the integral in the denominator. First observe that the expression in the denominator is not a function of $\theta$; it is a normalizing constant independent of $\theta$, and for this reason we can write $$ \pi \left( \theta | x \right)  \propto  p \left( x | \theta \right) \pi \left( \theta \right). $$  Eliminating all multiplicative constants in $p \left( x | \theta \right) \pi \left( \theta \right)$ that do not depend on $\theta$, we get that  $$\pi \left( \theta | x \right) = c  \cdot \theta^{x + \alpha - 1} \left( 1 - \theta \right)^{n-x + \beta -1},$$ where $c$ is some constant such that $$\displaystyle \int_0^1 c  \cdot \theta^{x + \alpha - 1} \left( 1 - \theta \right)^{n-x + \beta -1} d \theta = 1.$$The posterior $\pi \left( \theta | x \right)$ clearly takes the form of a beta distribution, making $$c = {\frac{\Gamma \left( \alpha +  \beta + n \right)}{\Gamma(\alpha + x) \Gamma( \beta + n - x)}}.$$ The posterior of $\theta$ is thus a beta distribution with parameters $\alpha^{\rm pst}$ and $\beta^{\rm pst}$ where \begin{eqnarray*} \alpha^{\rm pst} & = & \alpha + x, ~~~{\rm and}\\ \beta^{\rm pst} & = & \beta + n -x. \end{eqnarray*}

\begin{figure}[H]
   \begin{center}
      \includegraphics[height=3in, width = 3in]{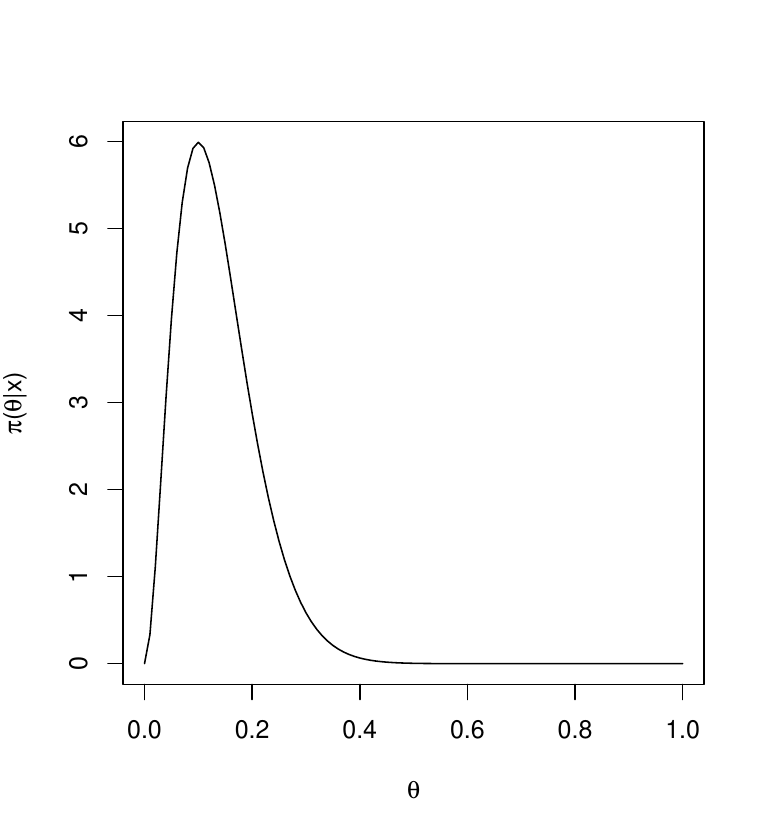}
   \caption[example] 
   { \label{fig:LowProbabilityPosterior} The posterior distribution with $n = 10$, $x=1$,  $\alpha = 2$ and $\beta = 10$.}
   \end{center}
\end{figure}

\begin{figure}[H]
   \begin{center}
      \includegraphics[height=3in, width = 3in]{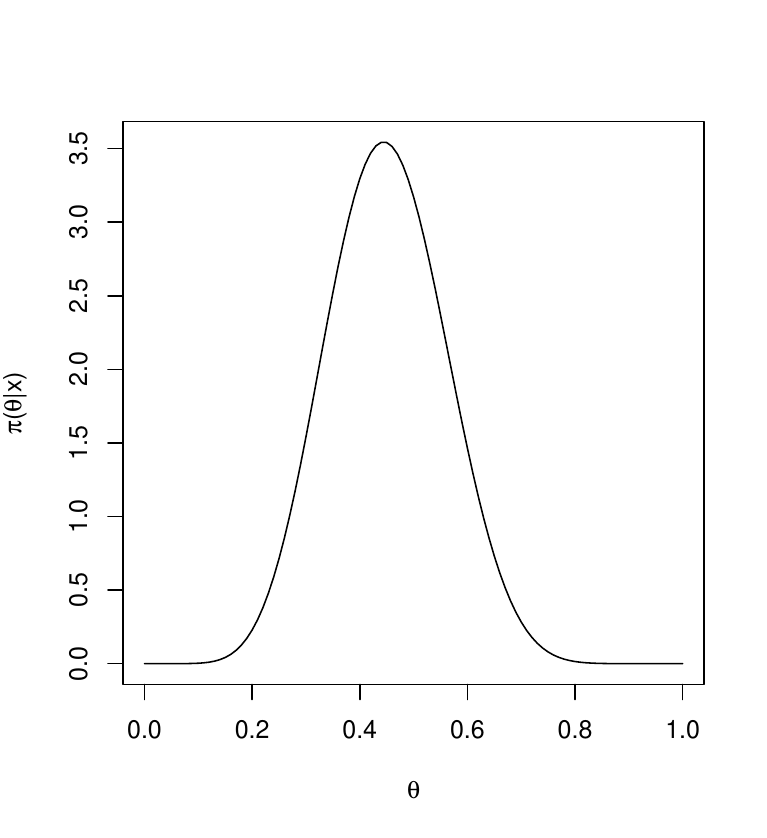}
   \caption[example] 
   { \label{fig:HighProbabilityPosterior} The posterior distribution with $n = 10$, $x = 2$,   $\alpha = 7$ and $\beta = 3$.}
   \end{center}
\end{figure}

The plots in Figures \ref{fig:LowProbabilityPosterior} and \ref{fig:HighProbabilityPosterior} show the posteriors corresponding to the priors shown in Figures \ref{fig:LowProbabilityPrior} and \ref{fig:HighProbabilityPrior}, respectively. In Figure \ref{fig:LowProbabilityPosterior} one success was observed after ten trials, emphasizing even more that the value of $\theta$ is small. Observe how the posterior in this case is more peaked at small values of $\theta$ than the prior was. In Figure \ref{fig:HighProbabilityPosterior} two successes were observed in ten trials, indicating that the probability of success was much smaller than the prior anticipated. Observe how, in this case, the peak of the posterior has significantly shifted to smaller values of $\theta$.

\subsection{The Prior and Posterior of the Entire System}\label{sctn:PostEntireSystem}

We now put this problem in the context of one large system that is composed of several subsystems. If all of the subsystems have to work for the entire system to work, how do the posterior distributions of the subsystem reliabilities inform the distribution of the total system's reliability? And how would testing the entire system (as a whole) affect the posterior of the total system reliability? The next subsection addresses the first question, and the second question is addressed in Section \ref{sctn:TotalSysTestSizing}.

\subsubsection{Subsystem Test Sizing}\label{subsctn:SubsystTestSizing}

This subsection focuses on how the distributions of the subsystem survival probabilities affect the distribution of the total system's survival probability.  If there are $S$ subsystems, and the entire system fails if any one of the subsystems fail (i.e., the subsystems work in series), then the success probability of the total system, $\theta_{\rm Tot~Sys}$, is  calculated as \begin{eqnarray} \nonumber  \theta_{\rm Tot~Sys} &  = & {\mathbb P} \left( {\textrm{Success of Subsys 1}} \right) \times  {\mathbb P} \left( {\textrm{Success of Subsys 2}} \right) \times \cdots  \\ & &  \times  {\mathbb P} \left( {\textrm{Success of Subsys}} ~S \right) \nonumber \\  \label{eqn:theta_ts} & = & \prod_{j=1}^S \theta_j,\end{eqnarray}  where $\theta_j$ is the success probability of the $j^{\rm th}$ subsystem. 

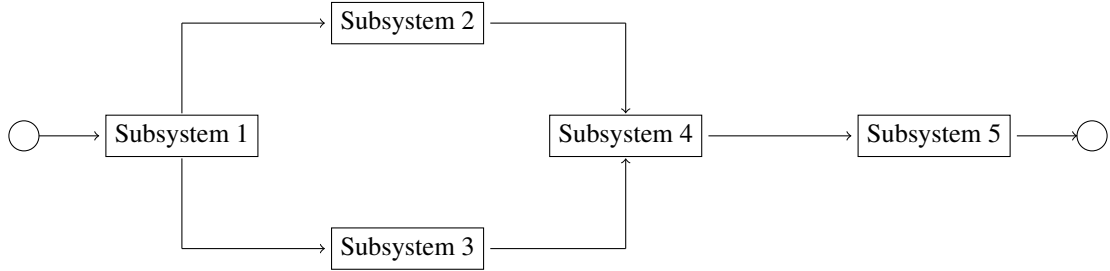
\begin{figure}[H]

\centering

\begin{tikzpicture}

\draw (2,2) circle (.2cm);
\draw [->] (2.2,2) -- (3,2);
\node[rectangle,draw] (r) at (4.1,2) {Subsystem 1};
\draw [-] (4.1,2.3) -- (4.1,3.5);
\draw [-] (4.1,1.7) -- (4.1,.5);
\draw [->] (4.1, 3.5) -- (6,3.5);
\draw [->] (4.1,.5) -- (6,.5);
\node[rectangle,draw] (r) at (7.1,3.5) {Subsystem 2};
\node[rectangle,draw] (r) at (7.1,.5) {Subsystem 3};
\draw [-] (8.2,3.5) -- (10,3.5);
\draw [-] (8.2,.5) -- (10,.5);
\draw [->] (10, 3.5) -- (10,2.3);
\draw [->] (10,.5) -- (10,1.7);
\node[rectangle,draw] (r) at (10,2) {Subsystem 4};
\draw [->] (11.1,2) -- (13,2);
\node[rectangle,draw] (r) at (14.1,2) {Subsystem 5};
\draw [->] (15.2,2) -- (16,2);
\draw (16.2,2) circle (.2cm);

\end{tikzpicture}

\captionsetup{width=.8\linewidth}
\caption{Flow chart of system composed of five subsystems, two of which work in parallel.}

\label{fig:parallelSystem_wtf}
\end{figure}

If $S = 5$ and the block diagram of the subsystems is as shown in Figure \ref{fig:parallelSystem_wtf}, then the system fails if Subsystem 1, 4, 5, or both 2 and 3 fail. In this case, the success probability of the entire system would be calculated as   \begin{eqnarray*} \label{eqn:theta_ts_parallel} \theta_{\rm Tot~Sys} & = & {\mathbb P}  \left(  \textrm{Success of Subsys 1}  \right) \times {\mathbb P} \left(  \textrm{Success of Subsys 2 or 3} \right) \times {\mathbb P} \left( \textrm{Success of Subsys 4} \right) \times {\mathbb P} \left(  \textrm{Success of Subsys 5} \right) \\ & = & \theta_1 \left( \theta_2 + \theta_3 - \theta_2 \theta_3 \right) \theta_4 \theta_5. \end{eqnarray*}

The value of $\theta_{\rm Tot~Sys}$ is thus the product and/or sum of beta random variables. The distribution of such a random variable has been derived in a number of publications, but this distribution is very complicated and thus difficult to work with analytically (see \cite{Fan},  \cite{Martz1}, \cite{Martz2}, \cite{Springer}, and \cite{Tang}). The distribution of $\theta_{\rm Tot~Sys}$ is easy to work with and understand, however, using Monte Carlo methods.

 Since the posterior distribution of all the components of the system take the form of a beta distribution with known parameters, assuming independence of the subsystems, we can easily simulate $n_{\rm Sim}$ values from the prior distribution of  $\theta_{\rm Tot~Sys}$. This requires simulating $n_{\rm Sim}$ $S$-tuples of $ \left( \theta_1, \theta_2, \theta_3, \ldots, \theta_S \right)$. With each simulated $S$-tuple, we can calculate a value of $\theta_{\rm Tot~Sys}$. The algorithm for generating $n_{\rm Sim}$ values of $\theta_{\rm Tot~Sys}$ for a system in series is given in Procedure 1; that for a system as shown in Figure \ref{fig:parallelSystem_wtf} is given in Procedure 2.

 \begin{algorithm}[ht]
 \SetKwInOut{Input}{input}
 \SetKwInOut{Output}{output}
 
 \hspace{-.4cm}{{\bf Procedure 1:} Simulating $n_{\rm Sim}$ values of $\theta_{\rm Tot~Sys}$ when the subsystems work in series}
 
 \BlankLine
 
 \Input{$\pi \left( \theta_1 | x_1 \right), \pi \left( \theta_2 | x_2 \right), \ldots, \pi \left( \theta_S | x _S\right), ~{\rm and}~ n_{\rm Sim}$, where $x_i$ is the number of successes of subsystem $i$}
 
 \Output{$\theta_{\rm Tot~Sys}^{(1)}, \theta_{\rm Tot~Sys}^{(2)}, \ldots, \theta_{\rm Tot~Sys}^{(n_{\rm Sim})}$}
 
 \For{$i\gets1$ \KwTo $n_{\rm Sim}$} {\
 
 	$\theta_{\rm Tot~Sys}^{(i)} \gets 1$\
	
	\For{$j \gets 1$ \KwTo $S$} {\
		
		Generate $\theta_j^{(i)} \sim \pi \left( \theta_j | x_j \right)$
		
		$\theta_{\rm Tot~Sys}^{(i)} \gets \theta_{\rm Tot~Sys}^{(i)} \cdot \theta_j^{(i)}.$
	
	}
 
 }

\end{algorithm}

 \begin{algorithm}[ht]
 
 \SetKwInOut{Input}{input}
 \SetKwInOut{Output}{output}
 
 \hspace{-.4cm}{{\bf Procedure 2:} Simulating $n_{\rm Sim}$ values of $\theta_{\rm Tot~Sys}$ when the five subsystems work as shown in Figure \ref{fig:parallelSystem_wtf} (series and parallel)}
 
 \BlankLine
 
 \Input{$\pi \left( \theta_1 | x_1 \right), \pi \left( \theta_2 | x_2 \right), \ldots, \pi \left( \theta_S | x _S\right), ~{\rm and}~ n_{\rm Sim}$, where $x_i$ is the number of successes of subsystem $i$}
 
 \Output{$\theta_{\rm Tot~Sys}^{(1)}, \theta_{\rm Tot~Sys}^{(2)}, \ldots, \theta_{\rm Tot~Sys}^{(n_{\rm Sim})}$}
 
 \For{$i\gets1$ \KwTo $n_{\rm Sim}$} {
 
 	$\theta_{2}^{(i)} \sim \pi \left( \theta_2 | x_2 \right)$
	
	$\theta_{3}^{(i)} \sim \pi \left( \theta_3 | x_3 \right)$
	
	$\theta_{2|3}^{(i)} \gets \theta_2^{(i)} +  \theta_3^{(i)} -  \theta_2^{(i)} \cdot  \theta_3^{(i)}$
 
 	$\theta_{\rm Tot~Sys}^{(i)} \gets  \theta_{2|3}^{(i)}$
	
	\For{$j \in \left \{ 1, 4, 5 \right \} $} {
		
		Generate $\theta_j^{(i)} \sim \pi \left( \theta_j | x_j \right)$
		
		$\theta_{\rm Tot~Sys}^{(i)} \gets \theta_{\rm Tot~Sys}^{(i)} \cdot \theta_j^{(i)}.$
	
	}
 
 }

\end{algorithm}

\begin{figure}[H]
   \begin{center}
      \includegraphics[height=3in, width = 3in]{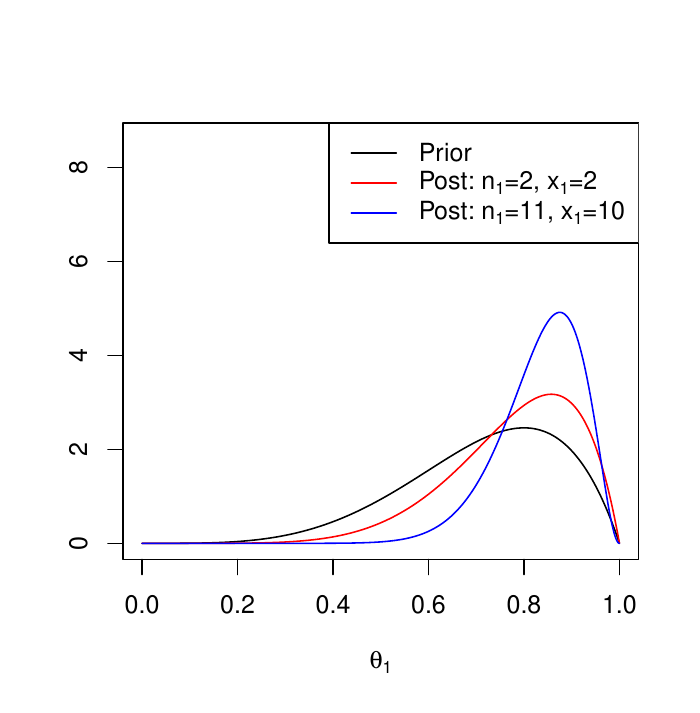}
   \caption[example] 
   { \label{fig:subsystExmpl1} The prior and posterior of the first subsystem with $\alpha = 5,$ and $\beta = 2$.}
   \end{center}
\end{figure}

\begin{figure}[H]
   \begin{center}
      \includegraphics[height=3in, width = 3in]{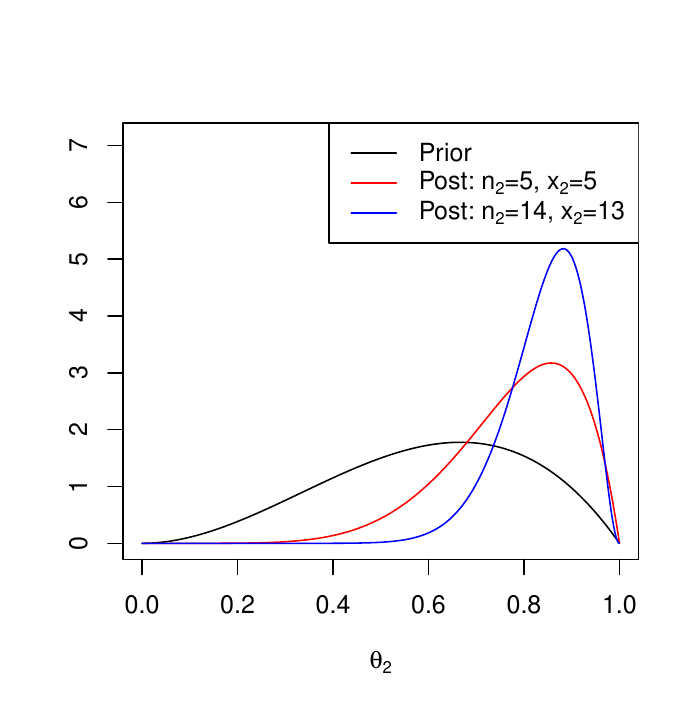}
   \caption[example] 
   { \label{fig:subsystExmpl2} The prior and posterior of the second subsystem with $\alpha = 3,$ and $\beta = 2$.}
   \end{center}
\end{figure}

\begin{figure}[H]
   \begin{center}
      \includegraphics[height=3in, width = 3in]{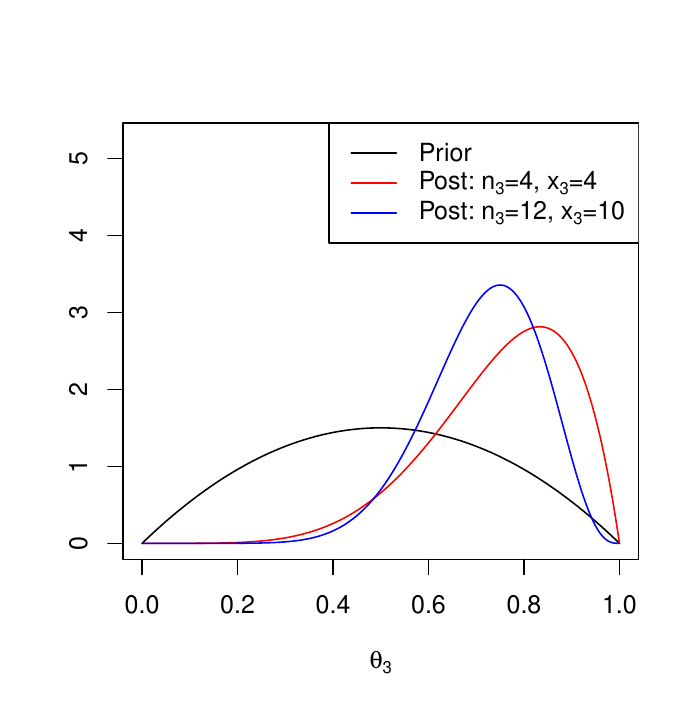}
   \caption[example] 
   { \label{fig:subsystExmpl3} The prior and posterior of the third subsystem with $\alpha = 2,$ and  $\beta = 2$.}
   \end{center}
\end{figure}

\begin{figure}[H]
   \begin{center}
      \includegraphics[height=3in, width = 3in]{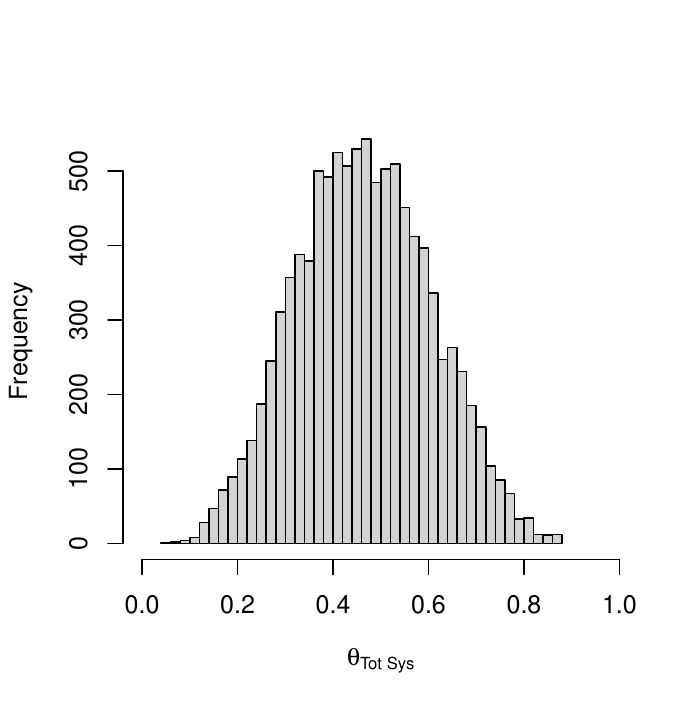}
   \caption[example] 
   { \label{fig:totalRelExmpl_1} The resulting distribution of $\theta_{\rm Tot~Sys}$ when $n_1 = 2$, $n_2 = 5$, and $n_3 =  4.$}
   \end{center}
\end{figure}

\begin{figure}[H]
   \begin{center}
      \includegraphics[height=3in, width = 3in]{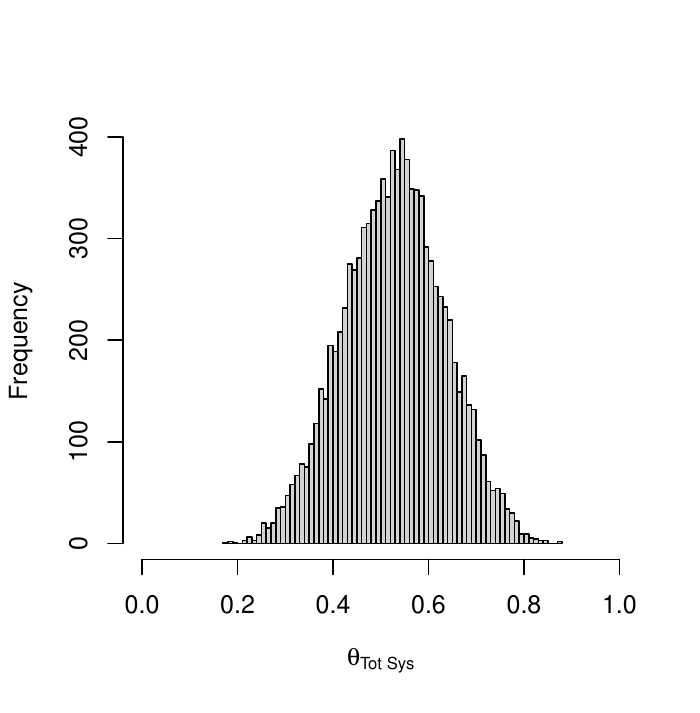}
   \caption[example] 
   { \label{fig:totalRelExmpl_2} The resulting distribution of $\theta_{\rm Tot~Sys}$ when $n_1 = 11$, $n_2 = 14$, and $n_3 =  12.$}
   \end{center}
\end{figure}

Figures \ref{fig:subsystExmpl1} $\textendash$ \ref{fig:ChangeInTotRelWIth7} illustrate how the posterior distributions of subsystem reliability affect the distribution of $\theta_{\rm Tot~Sys}$. In the simulations performed, we assumed the entire system was composed of three subsystems ($S = 3$) and that these subsystems worked in series. The priors of the three subsystems are shown in black in Figures \ref{fig:subsystExmpl1} $\textendash$ \ref{fig:subsystExmpl3}. The subsystems are then tested with $n_1 = 2$, $n_2 = 5$, and $n_3 = 4$, where $n_j$ is the number of times the $j^{\rm th}$ subsystem is tested. The resulting posteriors are shown in red in Figures \ref{fig:subsystExmpl1} $\textendash$  \ref{fig:subsystExmpl3}, and the resulting distribution of $\theta_{\rm Tot~Sys}$ is shown in Figure \ref{fig:totalRelExmpl_1}.\footnote{Unless otherwise indicated, all posterior samples shown in this paper have 10000 values.}  The subsystems were also tested at   $n_1 = 11$, $n_2 = 14$, and $n_3 = 12$, and the corresponding posteriors are shown in blue in Figures \ref{fig:subsystExmpl1} $\textendash$ \ref{fig:subsystExmpl3}.  Observe that these posteriors are more peaked (more informed) than the others since the sample sizes are larger. The distribution of $\theta_{\rm Tot~Sys}$ corresponding to these larger sample sizes is shown in Figure \ref{fig:totalRelExmpl_2}. Observe how the variance of this posterior is smaller than that shown in Figure \ref{fig:totalRelExmpl_1}; it should be since the subsystem sample sizes are larger.

\subsubsection{Total System Test Sizing}\label{sctn:TotalSysTestSizing}

We now investigate how testing the entire system (and not just its individual components) affects the posterior distribution of $\theta_{\rm Tot~Sys}$. Updating the prior distribution of $\theta_{\rm Tot~Sys}$ given test results on the total system is more challenging than updating the subcomponent values of $\theta$ because, in this case, the original distribution of  $\theta_{\rm Tot~Sys}$ is not a beta distribution. Recall that the prior distribution of $\theta_{\rm Tot~Sys}$ was analytically challenging to work with and, as a result, was obtained using Monte Carlo methods. It is not uncommon for practitioners to approximate this prior with another (perhaps Beta) distribution (see   \cite{Abdel}, \cite{Coelho}, \cite{Marques} and \cite{Tukey}) to make the posterior analysis simpler.    Others redefine the priors of the independent components entirely just so the prior and posterior of the total system's reliability is analytically tractable. Zoh et al. (see \cite{Zoh}), for example, set the priors of the components to negative log-gamma distributions.

{\bf These workarounds to make posterior analysis of the total system possible are not necessary!} In this paper we remind the reader of a simple and quick Monte Carlo method which updates the prior distribution of $\theta_{\rm Tot~Sys}$ after testing the entire system. This method  is a simple application of the algorithm proposed by Rubin (see \cite{Rubin}). Rubin observed that a sample from the posterior distribution of a parameter can be obtained by first generating values from its prior and then generating data conditioned on these sampled values. Those values of  the parameter for which the generated data match the observed data follow the posterior distribution.

To apply this algorithm in our case, we begin by writing the posterior for $\theta_{\rm Tot~Sys}$ as $$\pi \left( \theta_{\rm Tot~Sys}  | x_{\rm TS}  \right) \propto p \left( x_{\rm TS} | \theta_{\rm Tot~Sys} \right) \pi \left( \theta_{\rm Tot~Sys} \right),$$ where \begin{equation}\label{eqn:theLikelihood} p \left( x_{\rm TS} | \theta_{\rm Tot~Sys} \right) = {n_{\rm TS} \choose x_{\rm TS}} \theta_{\rm Tot~Sys}^{x_{\rm TS}} \left( 1 - \theta \right)^{n_{\rm TS}-x_{\rm TS}},\end{equation} $n_{\rm TS}$ and $x_{\rm TS}$ are the number of tests (and successes) of the total system.  Given that $x^*_{\rm TS}$ successes have been observed from $n_{\rm TS}$ trials of the total system, we sample from the posterior $\pi \left( \theta_{\rm Tot~Sys} | x_{\rm TS} = x^*_{\rm TS} \right)$ by first simulating from the prior of $\theta_{\rm Tot~Sys}$,  $\pi \left( \theta_{\rm Tot~Sys} \right)$.\footnote{An algorithm similar to the ones shown in Procedure 1 or Procedure 2 could be used to sample from the prior, $\pi \left( \theta_{\rm Tot~Sys} \right)$.} We then condition on these sampled values of $\theta_{\rm Tot~Sys}$ to generate candidate values of $x_{\rm TS}$ from the likelihood shown in Equation (\ref{eqn:theLikelihood}).  The simulated values of $\theta_{\rm Tot~Sys}$ for which the likelihood generates $x_{\rm TS} = x^*_{\rm TS}$ are then considered to be an exact sample from the posterior.  The details of this algorithm (assuming the subsystems work in series\footnote{Minor changes to the first `for' loop of the algorithm would be necessary if the subsystems did not work in series.}) are given below in Procedure 3, and the principles of this algorithm are identical to those illustrated in the example below.

 \begin{algorithm}[ht]
 
 \SetKwInOut{Input}{input}
 \SetKwInOut{Output}{output}
 
 \hspace{-.4cm}{{\bf Procedure 3:} Simulating $n_{\rm Sim}$ values of $\theta_{\rm Tot~Sys}$ from $\pi \left( \theta_{\rm Tot~Sys} | x_{\rm TS} = x_{\rm TS}^* \right)$}
 
 \BlankLine
 
 \Input{$\pi \left( \theta_1 | x_1 \right), \pi \left( \theta_2 | x_2 \right), \ldots, \pi \left( \theta_S | x_S \right), n_{\rm Sim}, n_{\rm TS},  x_{\rm TS}^*$}
 
 \Output{$\theta_{\rm Tot~Sys}^{(1)}, \theta_{\rm Tot~Sys}^{(2)}, \ldots, \theta_{\rm Tot~Sys}^{(n_{\rm Sim})}$}
 
 $i \gets 1$

 \While{$i < n_{\rm Sim}$} {
 
$ \theta_{\rm Tot~Sys}^{(i), ~{\rm cand}} \gets 1$

 \For{$j \gets 1$ \KwTo $S$} {
		
		Generate $\theta_j^{(i)} \sim \pi \left( \theta_j | x_j \right)$
		
		$\theta_{\rm Tot~Sys}^{(i), ~{\rm cand}} \gets \theta_{\rm Tot~Sys}^{(i),~{\rm cand}} \cdot \theta_j^{(i)}.$
	
	}
 
 Generate $x_{\rm TS}  \sim {\rm Binomial} \left( n_{\rm TS}, \theta_{\rm Tot~Sys}^{(i), ~{\rm cand}} \right).$
 
 \If{$x_{\rm TS} = x^*_{\rm TS}$} {

  $\theta_{\rm Tot~Sys}^{(i)} \gets \theta_{\rm Tot~Sys}^{(i),~ {\rm cand}}$

  $i \gets i + 1$
 
 }
 
 }

\end{algorithm}

\begin{tcolorbox}

\begin{center}

{\bf Example}

\end{center}

Consider two random variables, $Y$ and $Z$, each of which are defined on the integers 1, 2, 3, and 4.  Assume they follow the joint probability mass function $p(y,z)$ shown in Table 1.

\begin{center}
\begin{tabular} { |cc||c|c|c|c| }
\hline

 &  & \multicolumn{4}{c|}{$Z$} \\ 
 
   &  & \multicolumn{1}{c}{1} & \multicolumn{1}{c}{2} & \multicolumn{1}{c}{3} & \multicolumn{1}{c|}{4} \\ 
 
 
 \hline
 \hline

 \multirow{4}{1em}{$Y$} & 1 & .12 & .03 & .14 & .09 \\
 
  & 2 & .01 & .06 & .08  & .04 \\
  
  & 3 & .05 &  .09 &  .03 & .05 \\
  
  & 4 & .07 & .12 & .01 & .01 \\ 
  
  \hline

\end{tabular}

\captionof{table}{The joint probability mass function, $p(y, z)$.}
\end{center}

Now let us assume that we wish to generate values from the conditional distribution $p(y|z = 2)$. From the joint probability mass function given in Table 1, it is easy to calculate the conditional distribution of $Y$ given $Z = 2$, $$p(y|z = 2) = \left \{ \begin{array}{ll} .1 & y = 1 \\ .2 & y = 2 \\ .3 & y = 3 \\ .4 & y = 4 \end{array} \right. .$$   A straight-forward way to generate values from $p(y|z = 2)$ would be to first generate values of $Y$ from its marginal distribution, $p(y)$, where $$p(y) = \left \{ \begin{array}{ll} .38 & y = 1 \\ .19 & y = 2 \\  .22& y = 3 \\ .21 & y = 4 \end{array} \right. .$$ For each simulated value of $Y$, one could then generate a value of $Z$ from $p(z|y)$. Those values of $y$ for which the generated value of $Z$ is $2$ then form an exact sample from the conditional distribution of interest, $p(y|z = 2)$.

\end{tcolorbox}

\begin{figure}[H]
   \begin{center}
      \includegraphics[height=3in, width = 3in]{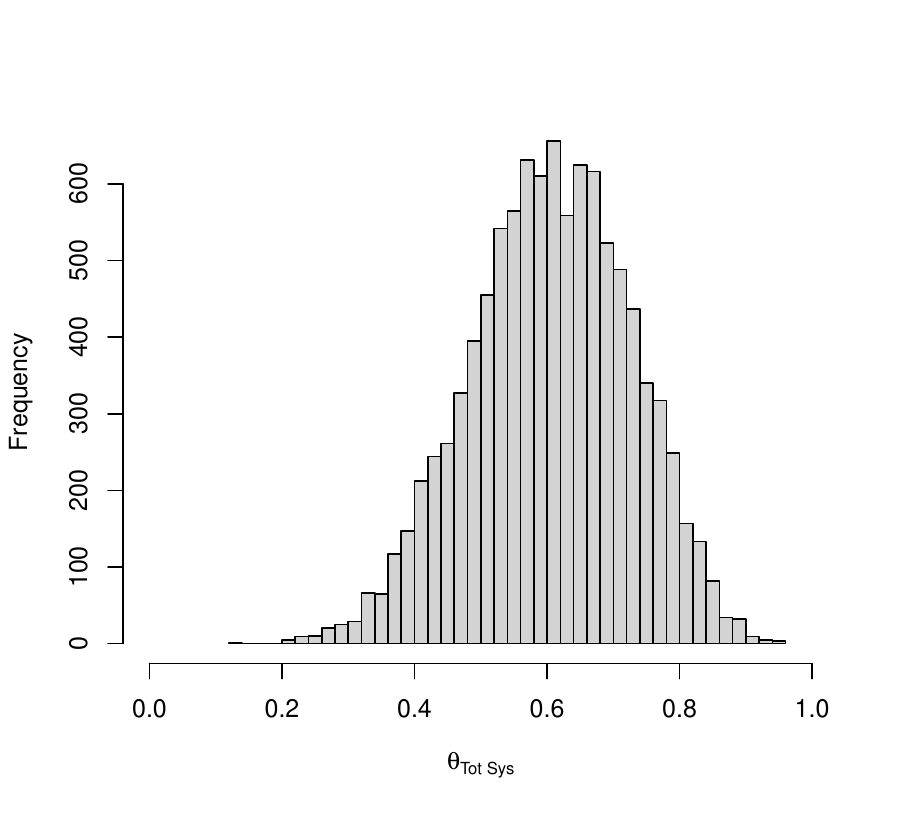}
   \caption[example] 
   { \label{fig:ChangeInTotRelWIth4} The resulting distribution of $\theta_{\rm Tot~Sys}$ when $n_1 = 2$, $n_2 = 5$,  $n_3 =  4,$ and $n_{\rm TS} = 4$.}
   \end{center}
\end{figure}

\begin{figure}[H]
   \begin{center}
      \includegraphics[height=3in, width = 3in]{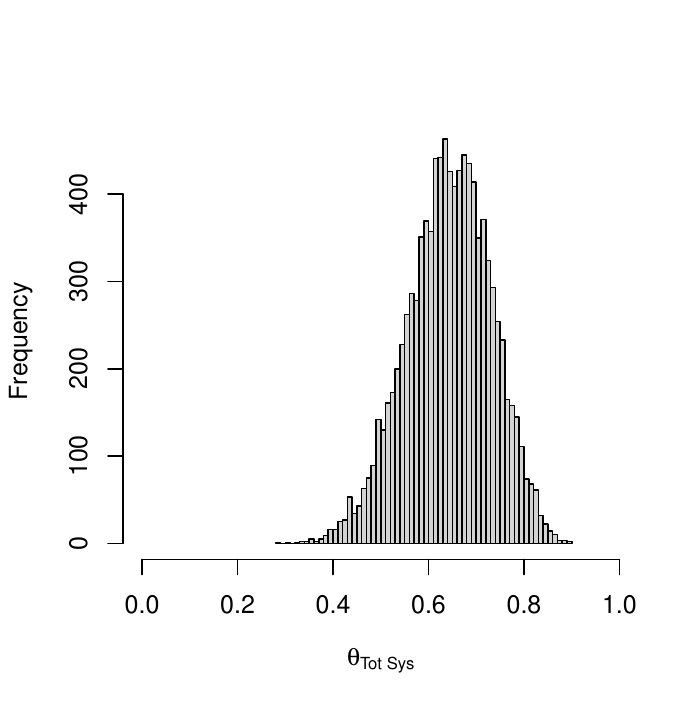}
   \caption[example] 
   { \label{fig:ChangeInTotRelWIth7} The resulting distribution of $\theta_{\rm Tot~Sys}$ when $n_1 = 11$, $n_2 = 14$,  $n_3 =  12,$ and $n_{\rm Tot} = 7$.}
   \end{center}
\end{figure}

The plots in Figures  \ref{fig:ChangeInTotRelWIth4} $\textendash$  \ref{fig:ChangeInTotRelWIth7} show how the distribution of $\theta_{\rm Tot~Sys}$ changes when tests on the entire system are executed. The plot in Figure \ref{fig:ChangeInTotRelWIth4} shows how the distribution of $\theta_{\rm Tot~Sys}$ changes from the distribution in Figure \ref{fig:totalRelExmpl_1} when 4 successes are observed out of four tests on the entire system. Observe that with this extra evidence of success, the distribution of $\theta_{\rm Tot~Sys}$ shifts to the right. The same story is told in Figure \ref{fig:ChangeInTotRelWIth7}. It shows how the distribution of $\theta_{\rm Tot~Sys}$ changes from the distribution in Figure \ref{fig:totalRelExmpl_2} when five successes are observed out of seven tests on the entire system. This distribution moves to the right and is also more peaked.

It is also critical that the reader understand the distinction between the two examples discussed above. In the first case (with the resulting distribution of $\theta_{\rm Tot~Sys}$ shown in Figure \ref{fig:ChangeInTotRelWIth4}), the total system was tested four times and four successes were observed. Because the system works in series, a successful test of the entire system implies a successful test of each component. The posterior distribution of $\theta_{\rm Tot~Sys}$ can thus easily be calculated  by simply updating the posteriors of the system's three components and then applying Procedure 1. This is not true for the second example (with the resulting distribution of $\theta_{\rm Tot~Sys}$ shown in Figure \ref{fig:ChangeInTotRelWIth7}). Recall that in the second example, the entire system was tested seven times, but only five successes were observed. Since it is not clear which component(s) failed (causing the failure of the entire system), the posteriors of the components cannot be updated, and Procedure 1 cannot be applied. In this case, Procedure 3 is necessary in calculating the posterior of $\theta_{\rm Tot~Sys}$.

\section{Conclusion}

This paper reviews some of the methodologies related to Bayesian reliability. It initially focuses on success/failure data of systems and their subsystems. The paper specifically addresses how the number of subsystem (or total system) tests affect the reliability of the entire system. It also reminds the reader of a simple  Monte Carlo method that can be employed to update the prior of a  total system's reliability when only data from the total system is available.

\end{document}